\documentclass{elsart}

\usepackage{natbib}

 \usepackage{epsfig}

\usepackage{amssymb}

\begin{document}

\begin{frontmatter}

% Title, authors and addresses

% use the thanksref command within \title, \author or \address for footnotes;
% use the corauthref command within \author for corresponding author footnotes;
% use the ead command for the email address,
% and the form \ead[url] for the home page:
% \title{Title\thanksref{label1}}
% \thanks[label1]{}
% \author{Name\corauthref{cor1}\thanksref{label2}}
% \ead{email address}
% \ead[url]{home page}
% \thanks[label2]{}
% \corauth[cor1]{}
% \address{Address\thanksref{label3}}
% \thanks[label3]{}

\title{A Computational Algebra Approach to the Reverse Engineering
of Gene Regulatory Networks}
 \author{Reinhard Laubenbacher\corauthref{cor1}}
 \ead{reinhard@vbi.vt.edu}
 \author{Brandilyn Stigler}
 \ead{bstigler@vbi.vt.edu}
 \address{Virginia Bioinformatics Institute at Virginia Tech\\
1880 Pratt Drive, Building XV, Blacksburg, VA 24061, USA}
 \corauth[cor1]{Corresponding author. Tel.: 540-231-7506; Fax.: 540-231-2606}

% use optional labels to link authors explicitly to addresses:
% \author[label1,label2]{}
% \address[label1]{}
% \address[label2]{}

%\author{Reinhard Laubenbacher}%, Brandilyn Stigler}
%\corauthor{Brandilyn Stigler}
%\ead{\{reinhard,bstigler\}@vbi.vt.edu}
%\address{Virginia Bioinformatics Institute at Virginia Tech\\
%1880 Pratt Drive, Blacksburg, VA 24061, USA}

\begin{abstract}
% Text of abstract
This paper proposes a new method to reverse engineer gene regulatory networks from 
experimental data.  The modeling framework used is time-discrete deterministic
dynamical systems, with a finite set of states for each of the variables.  The simplest
examples of such models are Boolean networks, in which variables have only two 
possible states.   The use of a larger number of possible states allows a finer discretization 
of experimental data and more than one possible mode of action for the variables, depending
on threshold values.  Furthermore, with a suitable choice of state set, one can employ
powerful tools from computational algebra, that underlie the reverse-engineering
algorithm, avoiding costly enumeration strategies.  To perform well, the algorithm requires 
wildtype together with perturbation time courses.  
This makes it suitable for small to meso-scale
networks rather than networks on a genome-wide scale.  The complexity of the algorithm
is quadratic in the number of variables and cubic in the number of time points.
The algorithm is validated on a recently published Boolean network model
of segment polarity development in {\it Drosophila melanogaster}.
\end{abstract}

\begin{keyword}
% keywords here, in the form: keyword \sep keyword
Reverse engineering \sep Gene regulatory networks \sep Discrete modeling 
\sep Computational algebra
% PACS codes here, in the form: \PACS code \sep code
\end{keyword}

\end{frontmatter}

%%%%%%%%%%%%%%%%%%%%%%%%%%%%%%%%%%%%%%%%%%%%%%%%%%%%%%%%%%
% main text
\section{Introduction}
%\label{intro}
Since the advent of DNA microarray technology several techniques from
mathematics, statistics, engineering, and computer science have been adapted or newly developed
for the purpose of using microarray and other data to reverse engineer the structure of
gene regulatory networks.  An important goal of systems biology is
to discover and model the causal relationships between the components of such networks
and the mechanisms that govern their dynamics (Kitano, 2002). 
Existing reverse-engineering methods can be broadly categorized as continuous vs. discrete and deterministic vs. 
stochastic.  Some methods aim to discover only the network topology, that is, which genes
regulate which others, with a directed graph or ``wiring diagram'' as output, 
possibly signed to indicate activation or inhibition.
The goal of other methods is to describe the dynamics of the network.
We first describe some of these methods in order to place the one proposed
in this paper into context.   

Several methods have been
proposed that reconstruct only the wiring diagram of the network. Bayesian network methods, first proposed 
in (Friedman \textit{et al.}, 2000), and further developed in (Hartemink \textit{et al.}, 2001), 
can be applied to single time points.  This approach was applied in (Hartemink \textit{et al.}, 2001) 
to data from 52 {\it Saccharomyces cerevisiae}
Affymetrix chips, in order to analyze and score models of the
gene regulatory network responsible for the control of genes necessary
for galactose metabolism.  The network considered involved 7 nodes.
In (Filkov \textit{et al.}, 2002) another statistical method was proposed that compares expression profiles from a time
series of measurements.  As an application the authors analyzed the data sets on {\it S. cerevisiae} in
(Cho \textit{et al.}, 1998) and (Spellman \textit{et al.}, 1998), which consist of four microarray time series.

A method adapted from metabolic control analysis was proposed in (de la Fuente \textit{et al.}, 2002).  It requires
input data based on very small perturbations of the rate of expression of each gene.  
The method was applied to simulated data using a model, published in (Mendoza \textit{et al.}, 1999), 
of the gene network
that controls flower morphogenesis in {\it Arabidopsis thaliana}, involving 10 genes.  
Using 10\% perturbations, the authors
were able to completely reconstruct the network of regulatory relations.  While this survey of methods
to reverse engineer gene networks in the form of a wiring diagram is far from complete, it provides a 
flavor of the variety of approaches, the scale at which the methods are being applied, and the
challenges that result from a lack of appropriate experimental data.

The most common approach to the modeling of dynamics
is to view a gene regulatory network as a biochemical network of gene products,
typically mRNA and proteins, and to describe their rates of change through a system of ordinary differential
equations.  Thus, the modeling framework is that of a time-continuous, deterministic dynamical system.   We
describe in detail the method in (Yeung \textit{et al.}, 2002), 
as it shares many features with our method, even though it uses
a different modeling framework.  According to the authors the method is intended to generate a
``first draft of the topology of the entire network, on which further, more local, analysis can be based.''
The authors make two assumptions.  

Firstly, the system is assumed to be operating near a steady state, so that the dynamics
can be approximated by a linear system of ordinary differential equations:
$$
\frac{dx_i}{dt}=-\lambda_ix_i(t)+\sum_{j=1}^Nw_{ij}(t)+b_i(t)+\xi_i(t),
$$
for $i=1,\ldots ,N$.  Here, $x_1,\ldots ,x_N$ are mRNA concentrations, the $\lambda_i$ are the self-degradation
rates, the $b_i$ are the external stimuli, and the $\xi_i$ represent noise.  The (unknown) $w_{ij}$ 
describe the type and strength of the influence of the $j$th gene on the $i$th gene.  They assemble to a 
square matrix $W$ of real numbers.  The output of the reverse-engineering algorithm is this matrix $W$.
The input is a series of data points obtained by applying the stimulus $(b_1,\ldots ,b_N)^T$
and measuring the concentrations $x_1, \ldots ,x_N$ $M$ times.  Assembling these measurements
into a matrix $X$, neglecting noise, and absorbing self-degradation into the coupling constants $w_{ij}$,
we obtain a matrix equation
$$
\frac{d}{dt}X=WX+B.
$$
Here, $X$ is an $(N\times M)$-matrix, $W$ an $(N\times N)$-matrix, and $B$ an $(N\times M)$-matrix.
Using singular value decomposition (SVD) one obtains
$$
X^T=UWV^T,
$$
where $U$ and $V$ are orthogonal to each other.  The first step is to obtain a particular
solution $W_0$ to the reverse-engineering problem.   One then obtains all possible solutions to the problem as
$$
W=W_0+CV^T,
$$
where $C$ ranges over the space
of all square $(N\times N)$-matrices whose entries are equal to 0 for a certain range of $j$ and
arbitrary otherwise.  Equivalently, $CV^T$ ranges over all matrices that vanish on the given
time points.

The second assumption made in the paper is that gene regulatory networks are sparse.  This provides
a selection criterion on which to base a particular choice of $C$, and hence of $W$.  That is, the
method selects the sparsest connection matrix $W$.  This is accomplished through a particular choice
of norm, and robust regression.  The algorithm was validated by way of simulated data from three
networks. 

The other end of the model spectrum takes the view of a gene regulatory network as a logical switching
network.  First proposed in (Kauffman, 1969), Boolean network models have the advantage of being
more intuitive than ODE models and might be considered as a coarse-grained approximation
to the ``real'' network.  They differ from ODE models in that time is taken as discrete and
gene expression is discretized into only two quantitative states, as either present or absent.  
There is increasing evidence that certain types of gene regulatory networks have key features that
can be represented well through discrete models or hybrid models;
see, e.g., (Filkov and Istrail, 2002).  Several algorithms for reverse engineering of Boolean network models for gene expression
have been proposed.  
The algorithm REVEAL (Liang \textit{et al.}, 1998) uses as a criterion for model selection the concept of so-called
mutual information.   For a given experimental data set (assumed to be already discretized into
a binary data set), that is, a given set of state transitions, the algorithm finds a Boolean function
for each node of the network that ``optimally'' determines the output from the input by using
as few variables as possible.  In essence,
the algorithm finds the sparsest possible Boolean network that is consistent with the data.
The search is done by enumeration.  In order to make this process feasible, the number of inputs
to the functions is restricted to at most three.  The algorithm has been tested by the
authors on simulated data
sets and was found to perform very efficiently.  Another algorithm for Boolean network identification
was given in (Akutsu \textit{et al.}, 1999).  It is in essence also 
an enumeration algorithm, applied to networks in which the
in-degree of each node is at most $2$.  

One of the disadvantages of the Boolean network modeling framework is the need to discretize
real-valued expression data into an ON/OFF scheme, which loses a large amount of information.
In response to this deficiency, multi-state discrete models and hybrid models have been developed.
The most complex one (Thomas, 1991; Thieffry \textit{et al.}, 1995; Thieffry and Thomas, 1998) 
uses multiple states for the genes in the network corresponding
to certain thresholds of gene expression that make multiple gene actions possible.  The authors
are most interested in the modeling and function of feedback loops.  The model includes a mixture
of multi-valued logical and real-valued variables, as well as the possibility of asynchronous
updating of the variables.  While this modeling framework is capable of better capturing
the many characteristics of gene regulatory networks than Boolean networks, it also introduces
substantially more computational complications from a reverse-engineering point of view, 
especially the ability  of asynchronous update, which adds orders of magnitude to the combinatorial
explosion of possibilities.  Multiple discrete expression levels were also used in the
reverse-engineering method in (Repsilber \textit{et al.}, 2002), 
which uses genetic algorithms to explore the parameter space
of multistage discrete genetic network models.  

In (Brazma and Schlitt, 2003) a hybrid modeling framework was introduced that tries to capture discrete as well
as continuous aspects of gene regulation.  The authors' finite state linear model has a Boolean
network type control component, as well as linear functions that represent an 
environment of substances that change their concentrations continuously.  According to the authors,
this framework can be generalized to more than two states for the logical variables.  
The reverse-engineering algorithm described in (Brazma and Schlitt, 2003) is based
essentially on enumeration of all possible
functions that fit a given data set.  

Finally, it is worth mentioning a Boolean network approach that incorporates stochastic features of
gene regulation.  Probabilistic Boolean networks have been
introduced in (Shmulevich \textit{et al.}, 2002).  
While this survey of existing reverse-engineering methods is by no means 
comprehensive, it provides a context for the new method proposed in this paper.  For
more thorough reviews the reader can consult, e.g., (Bower and Bolouri, 2001) or (de Jong, 2002).  

While continuous and discrete modeling approaches seem to be far apart, it is useful to keep in
mind that most ODE models cannot be solved analytically and that numerical solutions of
such systems involve the approximation of the time-continuous system by a time-discrete one. 
Furthermore, when validating an ODE model using microarray data it is often necessary
to utilize thresholds for continuous concentrations.  
The connection between an ODE system and an associated discrete system
that captures information about the continuous dynamics 
has been formalized in (Lewis and Glass, 1991).
So, in the end, the two modeling paradigms
might not be as far apart as it appears.  

The modeling framework we describe in the next section is of the discrete multi-state variety.
We propose here an approach that describes
a regulatory network as a time-discrete multi-state dynamical system, synchronously updated.
We further make the additional assumption that the set of states of the nodes in the network
can be endowed with the algebraic structure of a finite number system, which allows us to 
use techniques and algorithms from computational algebra.  
As mentioned earlier, it is difficult for any
reverse-engineering method to be validated using a real system as a test case, firstly
because a suitable data set might not be available and, secondly, model predictions might
be hard to verify without extra experiments.  For this reason, almost all methods discussed
above have used simulated data for validation.  We have chosen to use a recent
Boolean network model for segment polarity development in {\it Drosophila melanogaster}
(Albert and Othmer, 2003), consisting of sixteen nodes per cell, representing genes and gene products.  
The model is sufficiently complex to be of interest, but small enough so we can
compare in detail the agreements and disagreements of our reverse-engineered
network with the real one, thereby illustrating the performance and limitations
of our method.  

%%%%%%%%%%%%%%%%%%%%%%%%%%%%%%%%%%%%%%%%%%%%%%%%%%%%%%%%%%%%%%%%%%%%
\section{Reverse Engineering of Polynomial Systems over Finite Fields}

As mentioned above, we adopt the modeling framework of time-discrete multi-state dynamical systems.
Let $X$ be the set of possible states of the nodes of the network, and assume that $X$ is a 
finite (but possibly large) set.  One should think of $X$ as a set of discretized expression levels, 
for instance.  (We will address the issue of data discretization in a later section.)  Let 
$$
f:X^n\longrightarrow X^n
$$
be a discrete dynamical system over $X$ of dimension $n$.
Then $f$ can be described in terms of its coordinate functions
$
f_i:X^n\longrightarrow X,
$
for $i=1,\ldots ,n$.  That is, if ${\bf x}=(x_1,\ldots ,x_n)\in X^n$ is a state, then
$
f({\bf x})=(f_1({\bf x}),\ldots ,f_n({\bf x}))$.
We refer to such systems as {\it finite dynamical systems}.

The reverse-engineering problem we are focusing on is one of model selection
and can be stated as follows.

Given one or more time series of state transitions, generated by a biological system
with $n$ varying quantities,
choose a finite dynamical system $f:X^n\longrightarrow X^n$, which fits the data and ``best describes'' the
biological system.  To be precise, we presume that we are given a set of
state transitions of the network, in the form of one or more
time series.  That is, we are given sequences of states
$$
\begin{array}{lllll}
{\bf s}_1&=&(s_{11},s_{21},\ldots ,s_{n1}),&\ldots &,{\bf s}_m=(s_{1m},\ldots ,s_{nm})\\
{\bf t}_1&=&(t_{11},t_{21},\ldots ,t_{n1}),&\ldots &,{\bf t}_r=(t_{1r},\ldots ,t_{nr})\\
&&&\ldots&
\end{array}
$$
These satisfy the property that,
if the unknown transition function
of the network is $f$, then
\begin{eqnarray*}
f({\bf s}_i)&=&{\bf s}_{i+1}, \quad i=1,\ldots ,m-1\\
f({\bf t}_j)&=&{\bf t}_{j+1},\quad j=1,\ldots ,r-1\\
&\ldots&
\end{eqnarray*}
Typically, there will be more than one possible choice.  In fact, unless {\bf all} state transitions of the system
are specified, there will be more than one network that fits the given data set.  Since this much information
is hardly ever available in
practice, any reverse-engineering method has to choose from a large set of possible network
models.  In the absence of a good understanding of the properties and characteristics of gene regulatory
networks, one is limited to some type of Occam's razor principle for model selection.  Before describing
the principle we use, we first need to describe our computational framework.
 
We now make the further assumption that our state set $X$ is chosen so that it can be given the structure
of a {\it finite field} (Lidl and Niederreiter, 1997), that is, a finite number system.  
This is possible whenever the number of elements
in $X$ is a power of a prime number.  This can be easily accomplished by an appropriate choice of data
discretization.  One possible approach is to choose a prime number $p$ of possible
states, in which case the number system can be
taken to be $\mathbb Z/p$, the integers modulo $p$.  
An important consequence of this assumption is the well-known fact 
(see, e.g., (Lidl and Niederreiter, 1997), p. 369) that each of the coordinate functions of $f$ can be expressed
as a polynomial function in $n$ variables, with coefficients in $X$, and so that the degree of each variable
is at most equal to the number of elements in $X$.

{\bf Example.}
Boolean functions can be represented as polynomial functions with coefficients in 
$\mathbb Z/2={0,1}$.
Indeed, if $x$ and $y$ are Boolean variables, then
$$
x\wedge y=x \cdot y, \\
x \vee y=x+y+x \cdot y,\\
\lnot x=x+1.
$$

Note that addition does not correspond to the logical OR function, but rather the exclusive OR function,
XOR.  
Since every Boolean function can be written in terms of these three Boolean operations, we see that
every Boolean function can be represented as a polynomial function on the field with two elements.
Conversely, in light of the above fact, we see that any function 
$({\mathbb Z}/2)^n\longrightarrow ({\mathbb Z}/2)^n$ can be realized as a Boolean network.

So assume now that our state set is a finite field, denoted $k$.
As observed above, the model $f$ we are searching for is determined by its coordinate functions
$f_i:k^n\longrightarrow k$.  We can reverse engineer each coordinate function independently and 
thus reconstruct the network one node at a time.  Our algorithm is very similar in outline to that
in (Yeung \textit{et al.}, 2002).  We first compute the space of all networks that are consistent with the given time
series data.
We then choose a particular network $f=(f_1,\ldots ,f_n)$ that satisfies the following
property:

{\bf Criterion for model selection.}  
For each $i$, $f_i$ is minimal in the sense that there is no non-zero polynomial $g\in k[x_1,\ldots ,x_n]$
such that $f=h+g$ and $g$ is identically equal to zero on the given time points.  

That is, we exclude
terms in the polynomials $f_i$ that vanish identically on the data.   No reverse-engineering
method is able to detect such terms without prior information, even though they may exist in the real network but be 
nonfunctional under the particular experimental conditions.  Note that this criterion is different
from that used in (Yeung \textit{et al.}, 2002) and also by REVEAL.  In both of those algorithms the search is for the
sparsest network, that is, a network such that each node takes inputs from as few variables as
possible. 

In terms of the coordinate functions, the basic
mathematical reverse-engineering problem is then as follows, formulated for one time series.  The
statement of the problem for several time series is similar.

{\bf Problem.}  Suppose we are 
given a collection of states $\mathbf s_1=(s_{11},\ldots ,s_{n1}),\ldots ,\mathbf s_m=(s_{1m},\ldots ,s_{nm})$, and a 
choice of coordinate $i\in\{1,\ldots ,n\}$.
\begin{enumerate}
\item
Find all polynomial functions
$f_i\in k[x_1,\ldots ,x_n]$ such that 
$$
f_i(\mathbf s_j)=s_{i,j+1}.
$$
That is, find all polynomials that map $\mathbf s_j$ to the $i$th coordinate of the next time point $\mathbf s_{j+1}$.
\item  From that set of functions choose one that does not contain any terms that are identically equal to zero
at all time points (which is unique; see Appendix \ref{math}).
\end{enumerate}

The key advantage of the polynomial modeling framework over a finite field is that there is a well-developed
algorithmic theory that provides mathematical tools for the solution of this problem which have polynomial time
complexity in the input.  Thus, we can overcome one disadvantage that discrete models have compared to 
ODE models, for which there is a well developed mathematical theory.  This feature has been an important
reason for the use of polynomial systems over finite fields in engineering, in particular control theory
(see, e.g., (Marchand and Le Borgne, 1998)).  We now describe the algorithm.

We fix one coordinate/node in the network and reverse engineer its transition function.  To simplify
notation, we assume that we have given a time series $\mathbf s_1,\ldots ,\mathbf s_m$, as above, together with
elements $a_1,\ldots ,a_m\in k$, and we are looking for all polynomial functions $f\in k[x_1,\ldots ,x_n]$
such that $f(\mathbf s_j)=a_j$ for all $j=1,\ldots ,m$.
First we compute a particular polynomial $f_0$ that fits the data.  There are several methods to do this,
Lagrange interpolation being one of them.  We use the following formula, based on the so-called
Chinese Remainder Theorem (see, e.g., (Lang, 1971), p. 63):
%%%%%%%%%%%%%%%%%%%%%%%%%%%%%%%%%%%%
$$
f_0(\mathbf x)=\sum_{j=1}^ma_jr_j(\mathbf x),
$$
where the polynomials $r_j$ are defined as follows. Let $1\leq i\neq j<m$. If ${s}_{i}\neq {s}_{j}$, 
then find the first coordinate $l$ in which they differ. Define 
        $$b_{ij}(\mathbf x)=\left( s_{j,l}-s_{i,l}\right) ^{p-2}\left( x_{l}-s_{i,l}\right) $$
for every $i\neq j$. Then, for $i\neq j$,
        $$r_{j}(\mathbf x)=\Pi _{i=1}^{m-1}b_{ij}(\mathbf x).$$
If ${s}_{i}={s}_{j},$ then we have reached a limit
cycle. We can restrict the time series to the states
${s}_{1},\ldots ,{s}_{j-1}.$ 
We note that $r_k(\mathbf s_k)=1$ and $r_k(\mathbf x)=0$ otherwise.
%%%%%%%%%%%%%%%%%%%%%%%%%%%%%%%%%%%%%%%%%%%%%%%%
It is straightforward to check that this polynomial function does indeed interpolate the given time series.

Now consider two polynomials $f,g\in k[x_1,\ldots ,x_n]$ such that 
$$
f(\mathbf s_j)=a_j=g(\mathbf s_j).
$$
Then $(f-g)(\mathbf s_j)=0$ for all $j$.  That is, any two such functions differ by a function that
is identically equal to $0$ on the given time series.  So, in order to find all functions that fit the data,
we need to find all functions that vanish on the given time points.  Note that the set of all such functions
is closed under addition and multiplication by any polynomial function in $k[x_1,\ldots ,x_n]$, and
so forms a so-called {\it ideal} $I$ in the set of all polynomials.  
In Appendix \ref{math} we describe 
the details of an algorithm to compute
generators for $I$ (similar to a basis for a vector space).  The algorithm computes
polynomials $g_1,\ldots ,g_r\in I$ such that any $f\in I$ can be expressed as a linear combination
$$
f=\sum_{i=1}^rh_ig_i,
$$
for some polynomials $h_i\in k[x_1,\ldots ,x_n]$.  The next step is to reduce the polynomial $f_0$ found
in the previous step modulo the ideal $I$, that is, write $f_0$ as
$$
f_0=f+g,
$$
with $g\in I$.  Furthermore, $f$ is minimal in the sense that it cannot be further decomposed into $f'+h$
with $h\in I$.  In other words, $g$ represents the part of $f_0$ that lies in $I$, and that therefore is
identically equal to $0$ on the given time series.  Then we obtain all possible functions that
interpolate the time series in the form $f+g$, where $g$ runs through all elements of $I$.  
We summarize the different steps.

{\bf Reverse-engineering algorithm.}  (For one node of the network.) 

{\bf Input:} A time series $\mathbf s_1,\ldots ,\mathbf s_m\in k^n$ of network states, together with
expression levels $a_1,\ldots ,a_m\in k$.

{\bf Output:}  A polynomial function $f\in k[x_1,\ldots ,x_n]$ such that $f(\mathbf s_j)=a_j$ and such that
$f$ does not contain component polynomials that vanish on the time series.

\begin{enumerate}
\item
Compute a particular solution $f_0$ from the formula.
\item
Compute the ideal $I$ of all functions that vanish on the data.
\item
Compute the reduction $f$ of $f_0$ with respect to $I$.
\end{enumerate}

Some comments on the features of the algorithm and its complexity are in order.  Step (1) is straightforward.
Step (2) is where the bulk of the computation takes place.  The first observation is that the ideal $I$ can
be computed as the intersection of the ideals 
$$
I_j=\langle x_1-s_{j1},x_2-s_{j2},\ldots ,x_n-s_{jn}\rangle,
$$ 
for $j=1,\ldots ,m$.  Intersections of ideals can be computed algorithmically.  
See, e.g., (Cox \textit{et al.}, 1997), p. 185, for 
details.  The computation relies on the computation of a so-called Gr\"obner basis for several 
intermediate ideals.  It is known that the worst-case complexity of Gr\"obner basis calculations is
doubly exponential.  In the case of finite fields, however, the calculations needed can be reduced essentially
to linear algebra.  Our algorithm is very similar to one used in algebraic statistics, for the selection of
models for factorial design problems (Robbiano, 1998).  The only difference is that our algorithm computes $f_0$
and $I$ separately, whereas the algorithm in (Robbiano, 1998) combines the two.  (Since $I$ is the same for all
nodes, we only need to compute it once if done separately.)  It is shown there (Theorem 3.1) that
the algorithm is quadratic in the number of variables and cubic in the number of time points.  This
algorithm is implemented in the computer algebra system CoCoA ({\tt cocoa.dima.unige.it}).  For the results in this paper
we have used the computer algebra system Macaulay2 ({\tt www.math.uiuc.edu/Macaulay2}), which uses the standard algorithms
for computation of Gr\"obner bases.   We emphasize that no portion of this algorithm is based on enumeration.

As explained in Appendix \ref{math}, Gr\"obner basis calculations require the choice of an ordering of
the terms in the polynomials, in particular an ordering of the variables.  
This is necessary in order to carry out long division of multivariate
polynomials.  The end result of the calculations depends on the term ordering chosen, in the sense
that ``cheaper" variables, that is, those that are smallest in the ordering, will be used preferentially
in both the interpolation as well as the reduction calculations.  With prior information about the
existence of certain regulatory relationships this feature can be used to incorporate this prior 
knowledge into the algorithm.  (Likewise, using the Elimination Theorem (see Appendix \ref{math}), one can
exclude certain variables from appearing.)  In general, however, this dependence makes the 
interpretation of the precise form of the functions difficult at times.  In those cases we resort to building
a consensus model using a collection of variable orderings that makes each variable ``equally
costly on average.''  We can then extract from these different models common regulatory links and
common nonlinear terms.  In the next section we illustrate this strategy with a detailed discussion of an example.

%%%%%%%%%%%%%%%%%%%%%%%%%%%%%%%%%%%%%%%%%%%%%%%%%%%%%%%%%%%%%%%%%
\section{An Application}
\begin{figure}[h]
{\ \leavevmode
\hfil
\epsfysize=3.0truein\epsfbox{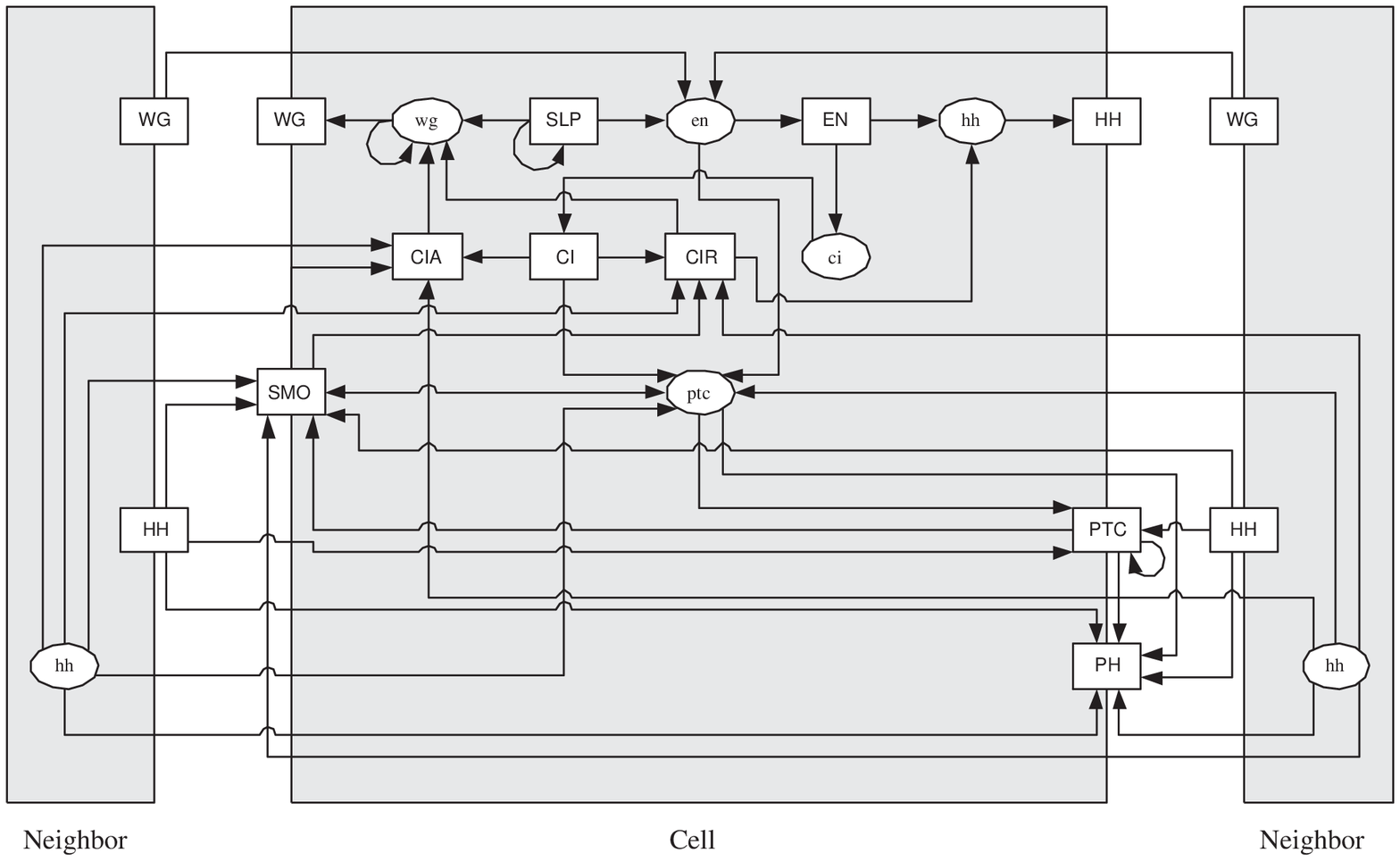} \hfil
}
\caption{The graph of interactions in the Boolean network developed in (Albert and Othmer, 2003). Ovals = mRNAs, rectangles = proteins. SLP denotes 2 forkhead domain transcription factors 
encoded by the gene \textit{sloppy paired} and is 
believed to activate the segment polarity genes depicted in the model (Cadigan \textit{et al.}, 1994). 
PH is the protein complex formed by the binding of HH to PTC (Ingham and McMahon, 2001). The protein SMO 
is encoded by the gene \textit{smoothened}. Because its transcription is not regulated by any molecular species in the model, \textit{smoothened} is not represented.} 
\label{bool}
\end{figure}

In this section, we present an application to the
well-characterized network of segment polarity genes in the
fruit fly \textit{D. melanogaster}. In (Albert and Othmer, 2003) and
(von Dassow \textit{et al.}, 2000) two models were suggested for the regulatory interactions in a
network of five segment polarity genes and their products. In (von Dassow \textit{et al.}, 2000) 
the authors proposed a continuous-state model using ordinary differential equations. 
In contrast, the Albert-Othmer model is discrete, in that the functions governing state transitions are 
Boolean functions. Both models were built from inferences given gene and protein expression data. Figure \ref{bool} depicts the graph of connections in the Boolean model in (Albert and Othmer, 2003).  In the graph, nodes represent mRNAs and proteins.  An edge between nodes indicates that the node at the 
head is involved in the regulation of the tail node. For example, an edge A$\to $B between protein nodes A and B implies that A regulates the 
transcription and translation of B, 
whereas an edge A$\to $\textit{b} from protein A to mRNA \textit{b} 
implies that A regulates the transcription of gene \textit{b}.  
Note that edges denote the existence of regulation, not the type, whether activation or inhibition.  
In Table \ref{sbf} which follows are some of the Boolean functions 
that accompany the model in Figure \ref{bool}. Superscripts denote time and subscripts location relative to 
the current cell. 

\begin{table}[h]
\begin{center}
\begin{tabular}{|l|}
\hline
$f_{6} =hh_{i}^{t+1}=EN_{i}^{t}\wedge \lnot CIR_{i}^{t}$ \tabularnewline
$f_{7} =HH_{i}^{t+1}=hh_{i}^{t}$ \tabularnewline
$f_{8} =ptc_{i}^{t+1}=CIA_{i}^{t+1}\wedge \lnot EN_{i}^{t+1}\wedge \lnot CIR_{i}^{t+1}$ \tabularnewline
$f_{9} =PTC_{i}^{t+1}=ptc_{i}^{t}\vee \left( PTC_{i}^{t}\wedge \lnot HH_{i-1}^{t}\wedge \lnot HH_{i+1}^{t}\right)$ \tabularnewline
\hline
\end{tabular}
\caption{Sample Boolean functions for the network in Figure \ref{bool}.}
\label{sbf}
\end{center}
\end{table}

The genes that are being modeled are \textit{wingless}
(\textit{wg}), \textit{engrailed} (\textit{en}),
\textit{hedgehog} (\textit{hh}), \textit{patched}
(\textit{ptc}), and \textit{cubitus interruptus} (\textit{ci}).
Our goal here is to
reverse engineer the Boolean network $\mathcal{N}$ in Figure \ref{bool} from 
time 
series generated by it, including null mutant time series for each of the five
genes in the network.

The network consists of a ring of 12 interconnected cells, which
are grouped into 3 parasegment primordia and the genes are
expressed every fourth cell. Since each cell of the ring has the
same network of segment polarity genes, we focus our work on the
reconstruction of the network in one cell, taking into account intercellular connections. For more details, see
(Albert and Othmer, 2003; von Dassow \textit{et al.}, 2000).  Note that for our purposes it
is irrelevant whether the model is indeed correct.

The starting point for our reverse-engineering algorithm is a collection of 
time series of discrete expression data.  We used known configurations of states 
(Cadigan \textit{et al.}, 1994; Hooper and Scott, 1992) for the 5 genes as initializations 
and generated times series for the wildtypes using the Boolean functions. 
All of the initializations terminate in steady states when evaluated by the Boolean functions in Table \ref{bf} in Appendix \ref{tables}. 
To account for intercellular dependencies we included 6 extra nodes into our model.  Table \ref{spf}
contains the translations of the Boolean functions in Table \ref{sbf} into polynomial functions in the
variables $x_1,\ldots ,x_{21}$, with coefficients in $\mathbb{Z}/2$.  (See Table \ref{pf} in Appendix \ref{tables} for all other polynomials.)

\begin{table}[h]
\begin{tabular}{|l l l|}
\hline
$f_{6}$ &=&$x_{5}\left( x_{15}+1\right)$  \\
$f_{7}$ &=&$x_{6}$ \\
$f_{8}$ &=&$x_{13}\left( X+x_{21}+ Xx_{21}\right)\left( x_{4}+1\right) 
\left( x_{13}\left( x_{11}+1\right) \left( x_{20}+1\right) \left( x_{21}+1\right) +1\right)$ \\
&&$X:=\left( x_{11}+x_{20}+x_{11}x_{20}\right)$ \\
$f_{9}$& =&$x_{8}+x_{9}\left( x_{18}+1\right) \left( x_{19}+1\right)+x_{8}x_{9}\left( x_{18}+1\right) \left( x_{19}+1\right)$ \\ 
\hline
\end{tabular}
\smallskip
\caption{Polynomial representations of the sample Boolean functions in Table \ref{sbf}. See Appendix \ref{tables} for the legend of variable names. To account for simultaneous updating we substituted simultaneously updated terms with their function expressions (e.g. in $f_8$ we replaced EN with \textit{en}).}
\label{spf}
\end{table}

The size of the state space of the Boolean network in each cell is
$2^{21}$, involving multiple components. Any single trajectory in that
space vastly underdetermines the network. 
If we use no known biological information other than the expression data
of the wildtype to construct a discrete polynomial model, our method predicts the
network to have 20 links, of which 14 are in the network ($\mathcal{N}$
has 44 links). (A description of the method with which we construct the consensus model is given below.)  The 
performance of the algorithm 
dramatically improves, however, if we incorporate
knock-out time series for the five genes, which we demonstrate below.

\begin{figure}[h]
{\ \leavevmode
\hfil
\epsfysize=3.0truein\epsfbox{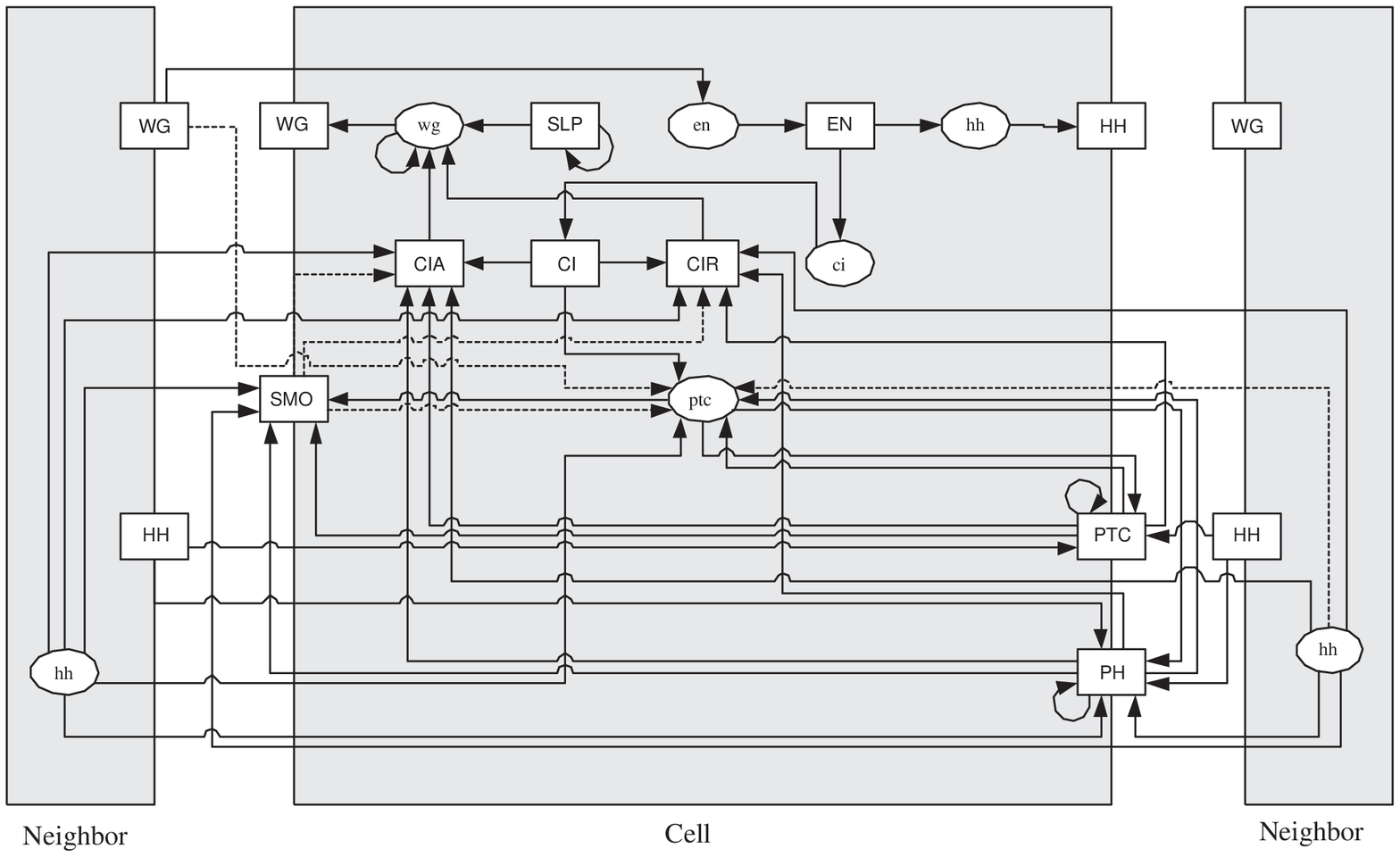} \hfil
}
\caption{The graph of the consensus model, constructed from the wildtype and knock-out time series.  The graph is the intersection of 
4 polynomial models, one for each variable ordering used. Dashed lines are links that appear in 3 of the 4 models.}
\label{common}
\end{figure}

To simulate an experiment in which node $x_i$ representing a gene is knocked out, 
we set its corresponding update function $f_i$ in Table \ref{pf} to 0 (and kept all other functions the same). 
We also set the corresponding functions in neighboring cells equal to 0.
We then generated a new time series by iteration of the given initializations. 
With these data, we constructed the model in Figure \ref{common}. 

Recall that we first 
construct all polynomial models (sets of polynomial functions) which fit 
a discrete time series. Then we select the one which is minimal with respect to summands in each
of the functions which evaluate to 0 on all time points. The algorithm relies on the choice of a total ordering for all possible monomials
in the given variables, in particular, a total ordering of the variables themselves.  The effect of
such a variable ordering is that the ``cheaper" variables, those that are smaller in this ordering,
are used preferentially in the algorithm.  In order to counteract this dependency, we use a 
consensus model extracted from four different choices of variable ordering: the
ordering $x_{1}<\cdots <x_{21}$, the reverse ordering $x_{1}>\cdots >x_{21}$, and 2 orderings which
make ``interior" variables greatest and least. Using each ordering
we ran the algorithm and intersected them to obtain a
consensus model (see Figure \ref{common}).

The consensus model we construct, which incorporates data from the wildtype and knock-out mutants, results in prediction of 37 of the 44 links
in the Boolean network.  We correctly identified the species which regulate the transcription of genes \textit{wg} and \textit{ci} and the synthesis of proteins SLP, WG, EN, HH, PTC,
and CI. We summarize the model of interactions and compare it
to the model in Figure \ref{bool}. In this model we assume that
transcription and translation occur in 1 consecutive time step. We will
discuss reconstruction of certain dynamic properties at the end of the
section.

In determining which species affect the transcription of the gene \textit{hh}, 
we found a function that involves fewer terms, than the polynomial representation of its 
counterpart in the Boolean model. Specifically, the function is in terms of the gene product EN. 
The function satisfies all time series, including knock-out data, generated by its associated Boolean function. 
The discrepancy lies in the
existence of a term in the Boolean function which does not contribute to the updating of \textit{hh}, that is, a 
term that is constant equal to $0$ on all input time points.
However, links whose effects are not reflected in the given data
are not detectable by any reverse-engineering method unless prior information about
the link is given.  
Similarly, the Boolean function for \textit{en} also contains such terms, 
which accounts for differences in regulation dependencies. 
We infer that transcription of the gene \textit{en} is regulated by Wingless protein in an anterior cell.

In this model, transcription of the \textit{patched} gene to synthesize \textit{ptc} mRNA in a given cell is regulated
by the following molecular species:  the proteins PTC, PH, SMO, CI and WG
in the cell anterior, as well as \textit{hh} from neighboring cells. We were not able to conclude that \textit{en} regulates
the transcription of the gene.  It is worth observing that this regulatory
link is not supported in the von Dassow model.  The Albert-Othmer model
predicts that transcription of the \textit{ptc} gene is regulated only by the whole
CI protein and its repressor form CIR.  In fact, the authors explain that
the cleavage of CI creates the repressor fragment CIR, whose synthesis
is repressed by \textit{hh} product in neighboring cells.  This effect may be too
indirect to be recovered from the data.  Our identification of WG protein in adjacent cells as a regulator and our failure to detect activation or repression by
\textit{engrailed} (whose transcription depends on WG) suggest that WG may
have a more direct effect on the regulation of \textit{ptc}. Further, it is
possible that existence of regulation by \textit{en} is
inconclusive, given the data, especially in light of the discrepancy
between the Boolean and continuous models. Currently we have no evidence
to support regulation by PTC or by PH.

For the protein complex PH, we detected that its synthesis is regulated itself and by \textit{ptc}, 
\textit{hh} and their corresponding products in adjacent cells. 
Recall that PH is the molecule formed when HH from adjacent cells binds to the transmembrane receptor PTC. 
In (Albert and Othmer, 2003) the authors assume in their model that this binding occurs 
instantaneously since it is known that the reaction occurs faster than transcription or translation 
(which they also presuppose to require 1 time unit for completion). Therefore, we infer that the synthesis of PH at a given time step is determined by the expression levels of \textit{ptc}, \textit{hh} and \textit{hh} product, from neighboring cells, in the previous time step. However, we did not identify PTC as a direct regulator. As PH may be transported into other cells via exocytosis (Taylor \textit{et al.}, 1993),  
its expression levels may not be maintained.  This suggests that its presence in a cell signals its transport out of the cell, hence implying a type of ``autoregulation''.

We identified the gene \textit{ptc}, its product, the complex PH, and
extracellular \textit{hh} to affect synthesis of SMO.  The binding of
PTC to SMO inactivates it via post-translational modification (Ingham,
1998).  However, if PTC binds to HH, then SMO is activated (Ingham, 1998).
So the translation of \textit{smoothened} mRNA is dependent on the synthesis of the complex PH.  (The mRNA \textit{smoothened} was not included in the model since the gene is transcribed ubiquitously throughout the segment (Ingham, 1998) and its transcription is not regulated by any biochemical in the model.)
Since this binding is assumed to occur instantaneously, as described above, we instead detect the regulation by 
\textit{hh} in adjacent cells and not its product.

While we correctly identified regulation by SMO, CI, and \textit{hh} in neighboring cells for the activating (CIA) 
and repressing (CIR) transcriptional factors transformed from CI, 
we predicted two other sources of regulation, namely by PTC and PH. 
There is evidence that PH activates signalling by SMO (Chen and Struhl, 1998), 
causing the transformation of CI into CIA, inhibiting cleavage of CI.  
Further, the rate of cleavage is a function of the amount of free PTC in the cell (von Dassow \textit{et al.}, 2000).
The presence of the links from PTC and PH to CIA and CIR suggests that their indirect regulation can be detected in the data.

To summarize the performance of our algorithm for detecting network topology, 
we show that the use of expression levels of wildtype and null
mutants enables detection of about 84\% of the interactions in 
the Boolean model. In contrast, using data only from the wildtype yields a mere 32\% identification.  

Next we focus on reverse engineering the dynamics of the Boolean model,
that is, the Boolean functions on each of the network nodes.  As pointed out
above, the functions in the Boolean model contain terms that evaluate to
$0$ on all input data, so that we are unable to detect the corresponding
relationships.  In order to compare the dynamics predicted by our method 
with the Boolean model
we reduce the polynomials in Table \ref{pf} by removing vanishing terms,
as described in the previous section. Note, however, that this reduction
depends on the choice of term ordering, as does the output of our algorithm.
For each choice of term ordering the reduced functions of the Boolean model
and the functions reverse engineered from the data agree exactly (see Table \ref{srbf}).  This
shows that we are able to completely predict the Boolean model's dynamics
from the wildtype and knock-out data.  However, due to the sensitivity of
our method to the chosen term ordering, the particular form of the reverse-engineered
functions may not be directly interpretable in terms of regulatory relationships.
We therefore proceed as before by extracting information about nonlinear
terms containing products of more than one variable
common to the reverse-engineered functions for multiple
term orderings.  These are analogous to mixed terms in ODE models.

In the consensus model constructed from the wildtype, all polynomials are
linear, suggesting that the concentrations of all molecular species in the
network are regulated by biochemicals working independently. However, Albert
\textit{et al.} (2003), von Dassow \textit{et al.} (2000), and others argue differently,
as indicated in their models. For a more accurate dynamics, we include expression data for the null mutants.

In the consensus model built with knock-out data, we found the following
nonlinear interactions. For the proteins SMO and PH, we found that their
synthesis depends on the ``interaction'' between the genes \textit{ptc}
and extracellular \textit{hh}. Specifically, the terms $[ptc]*[hh_{i-1}]$
and $[ptc]*[hh_{i+1}]$ appear in the polynomial functions for SMO and PH,
for all term orders,
which also appear in the Boolean functions for these two proteins. In the
function describing transcription of \textit{wg}, the terms $[wg]*$[CIA]
and $[wg]*$[CIR] are present, implying that its expression results from
interactions between itself and the activator and repressor forms of the
protein CI. These products also appear in the Boolean function for
\textit{wg}. Furthermore, the function of the protein fragments of CI
include the terms [PTC]$*[hh_{i-1}]$ and [PTC]$*[hh_{i+1}]$, both of which
exist in the corresponding Boolean functions. Finally, in any cell, our
method detects that PTC and \textit{hh} from a neighboring cell interact,
affecting the transcription of \textit{ptc}. As mentioned above, we have
no evidence to support this implication.

Our method shows a marked improvement when knock-out data are included. We 
were able to reconstruct about 84\% of the topology of the Boolean 
network, versus only 32\% when knock-out data are not included.  Further, 
we identified 10 nonlinear interactions, versus none in the model 
constructed with only wildtype data. 

\begin{table}[h]
\begin{tabular}{|l l l|}
\hline
$f_{6}$ &=&$x_{5}$ \\
$f_{7}$ &=&$x_{6}$ \\
$f_{8}$ &=&$x_{12}x_{13}+x_{13}x_{16}+x_{13}x_{20}+x_{18}x_{20}+x_{13}x_{21}+x_{19}x_{21}+x_{13}+x_{18}+x_{19}$ \\
$f_{9}$ &=&$x_{10}x_{14}+x_{14}x_{18}+x_{14}x_{19}+x_{9}x_{20}+x_{18}x_{20}+x_{9}x_{21}+x_{19}x_{21}+x_{8}+x_{9}+x_{10}$ \\
\hline
\end{tabular}
\smallskip
\caption{Sample Boolean functions reduced by the ideal of wildtype and knock-out time series.}
\label{srbf}
\end{table}

%%%%%%%%%%%%%%%%%%%%%%%%%%%%%%%%%%%%%%%%%%%
\section{Data Discretization}
\begin{figure}[h]
\begin{center}
\epsfxsize=2in
\epsfbox{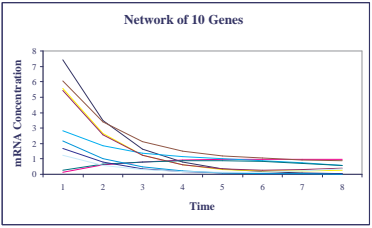}
\end{center}
\caption{The graph of the real-valued time series for a network $\mathcal{G}$ of 10 genes.} 
\label{net}
\end{figure}

\begin{figure}[h]
\begin{center}
\epsfxsize=2in
\epsfbox{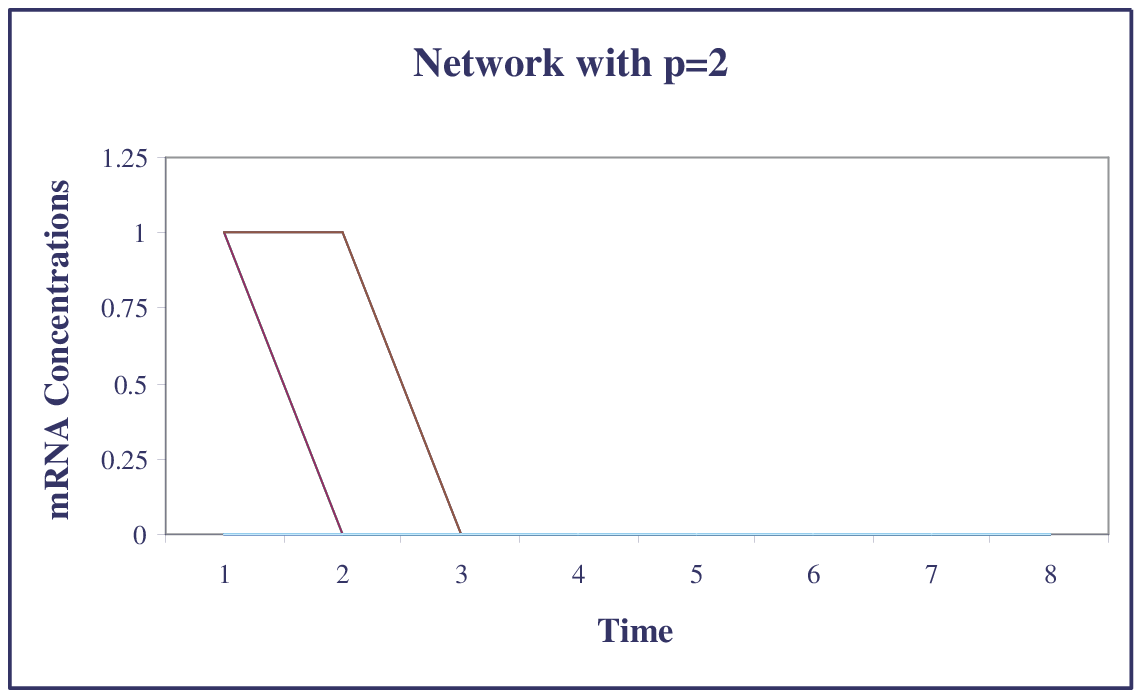}
\epsfxsize=2in
\epsfbox{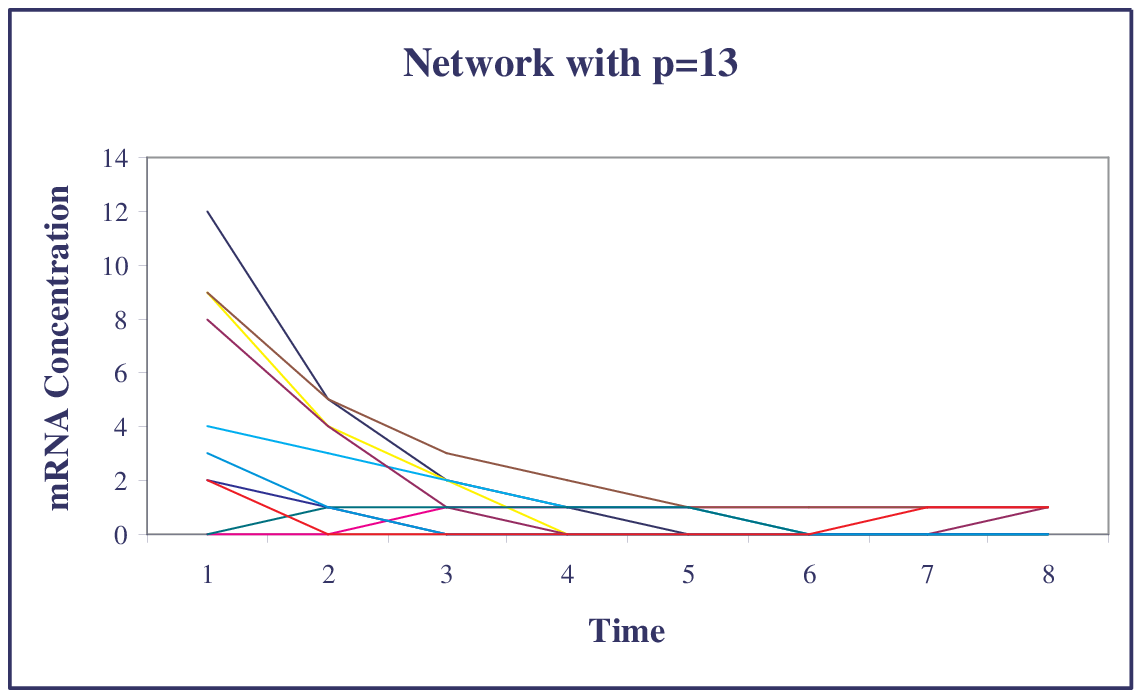}
\end{center}
\caption{The graph of the discrete time series for $\mathcal{G}$: $p=2$ (left) and $p=13$ (right) with discretization method = global as described above.}
\label{nets}
\end{figure}

Since gene expression data are real numbers, the first step in any reverse-engineering algorithm
using discrete models must be to discretize these real numbers into a finite (typically small)
set of possible states.  For Boolean networks this simply amounts to the choice of a single threshold
for the expression level of each gene, below which a gene is considered inactive.  The issue of data discretization
has been studied extensively, in particular from the points of view of Bayesian network applications
and machine learning; see, e.g., (Dougherty \textit{et al.}, 1995) and (Friedman and Goldszmidt, 1996).
Obviously, the way one discretizes the data plays an important role in what model one obtains. 
The first important choice is the number of discrete states allowed.  In our case, this is the choice
of $p$ in the algorithm.  It follows from results in (Green, 2003) that for $p$ large enough
the result of the reverse-engineering algorithm does not depend on $p$ anymore,
in the sense that the monomials in the polynomials remain the same, possibly with 
different coefficients.  While that paper
does not give an algorithm for choosing a suitable $p$, extensive experiments with networks simulated
with the biochemical network simulation program \textit{Gepasi} ({\tt http://www.gepasi.org})
suggest that for data sets up to 50 nodes an optimal $p$ is in the range between $5$ and
$13$.  The effect of  the choice of $p$ is demonstrated in Figures \ref{net} and \ref{nets}.
This example also demonstrates that the availability of more than two discrete states can be very
helpful for modeling purposes.  

%   Double file:

There are various ways to attach a discrete label to real-valued data. 
Thresholds with biological relevance is one type of labeling that can be used.  
For example, up-regulation, no regulation, and down-regulation of a gene are two thresholds that 
can partition a given data set into 3 groups, with labels $1,-1,0$, respectively.  
This maps the real-valued data into values in $\mathbb{Z}/3$.  
More thresholds can be integrated to refine the partitioning of the data set. 
Another method of discretization is to normalize the expression of each gene or protein and use
the deviation from the mean to discretize the data.  

%%%%%%%%%%%%%%%%%%%%%%%%%%%%%%%%%%%%%%%%%%
\section{Effect of Noise on the Algorithm}
In the previous section, one can see that the choice of prime provides different levels of 
resolution of the real-valued time series, yielding sharp differences in the discrete time series.  
To minimize this effect, a suitably large prime should be considered (Green, 2003).  
However, data collected from experimental processes typically have errors introduced due to deficiencies 
in methodology or imperfections in instrumentation, as is the case in microarray data.  
Furthermore, biological systems, such as biochemical pathways, exhibit variability in population 
or concentration levels.  For our purposes it is important to quantify this noise.

Because our method begins by discretizing real-valued time series, one would expect 
that discretization will smooth out some of the noise.  To test this hypothesis, 
we incorporated random noise, using a normal distribution, into time series generated from 100 simulated 
networks generated by the AGN software described in (Mendes \textit{et al.}, 2003).  
We generated 25 networks with 20 nodes for each of the 4 topologies supported by AGN.  
All time series are of length 11 time steps.  
We added noise in two ways: first, as a percentage of each data point, 
simulating biological variation, and second, as a percentage of a fixed value, 
simulating instrumentation error.  In both cases, we added 10\% and 25\% noise.  
We chose 3 different thresholds (0.5, 1.0, and 2.5) with which to Booleanize all time series.  
Then we computed the median number of entries 
that were changed in the noisy time series for both 10\% and 25\% noise, as compared to the original time series.  

The median number of entries that changed for all experiments is 5 (~2.3\% of 220).  In the case 
when noise is a percentage of each data
point, the median is 2
(~1\% of 220), when the threshold was restricted to 0.5 or 2.5. 
For the experiments described above, we infer that ~10\% of the noise added is propogated to the discrete time series.   

To test for the effect of noise on our method, we added 1\% noise to 
the discrete time series for the Boolean functions in (Albert and Othmer, 2003).  
(We estimate this to correspond to approximately 10\% noise in the ``real'' data, given the above results.)  
We then applied our method to the noisy time series and used 8 variable orderings 
to construct a consensus model.  For 6 of the nodes (SLP, \textit{wg}, WG, \textit{en}, HH, and CI) 
we were able to identify all of the correct links, plus some false positives.  
For 5 of the nodes (\textit{ptc}, PTC, SMO, CIR, and CIA), we identified half of the correct links.  
For the remaining nodes, we had no conclusive results.

%%%%%%%%%%%%%%%%%%%%%%%%%%%%%%%%%%%%%%%%%%
\section{Discussion}
We have described a new method to discover regulatory relationships between
the nodes in a gene regulatory network from data.  The modeling framework is that
of time-discrete dynamical systems with finite (but possibly large)
state sets for the variables.  The
crucial further assumption is that the set of possible states for a variable can
be given the structure of a finite field.
As a consequence, we
have available to us the very well-developed theory of algorithmic polynomial algebra, with
a variety of implemented procedures.  It is this machinery that we employ for the task
of reverse engineering.  In giving up a bit of freedom in the choice of state sets, we gain
a mathematical framework well suited for reverse-engineering purposes.  Our method is
very similar to that in (Yeung \textit{et al.}, 2002), in that we first compute all possible networks that
fit the given data.  We do this not by enumeration, but rather by a procedure similar
to describing all solutions of a homogeneous system of linear equations by computing
a basis for the nullspace.  Then, as in (Yeung \textit{et al.}, 2002), we choose a particular network.  Our 
selection criterion is to choose polynomial functions that do not contain any terms
which are identically equal to 0 on the given data set.  So, we choose a minimal
network, not in terms of network connections like in (Yeung \textit{et al.}, 2002) 
and (Liang \textit{et al.}, 1998), but rather
in terms of the structure of the functions.  The rationale is that if a term vanishes on
the data input, then no reverse-engineering method should be able to identify
it without prior biological knowledge.

We contrast our method with the discrete methods described in the introduction.
In its essence, the Thieffry-Thomas model can also be represented as a function from
a finite set of states to itself, with dynamics generated by iteration of this function.  
The lack of any further mathematical structure in the model makes its analysis very difficult.
Moreover, this modeling framework does not lend itself to mathematical reverse-engineering methods.
Even in the much simpler case of Boolean networks, all the algorithms  discussed above rely
at some point on enumeration of a large number of possible components of the
putative network.  

%Our method does not use enumeration at any point, using instead several algorithms
%from computational algebra.  The method
%has very good complexity, being quadratic in the number of variables and cubic in the number
%of time points.  Its main drawback is that it requires wildtype as well as some perturbation data
%in order to give good results.  Detailed studies of the
%effects of noise remain to be carried out.  Furthermore, a comprehensive study of the effect
%of different data discretizations needs to be completed.

For method validation we used the Boolean model of the {\it D. melanogaster}
segment polarity network recently published in (Albert and Othmer, 2003).  
We generated time series and perturbed time series from
this model and used them as input for the algorithm.
Note that for our purposes it is irrelevant whether the model is in
fact an accurate depiction of the real system.  Of the 44 links in the model we correctly
identify 37, together with 13 links not present in the model.  For each of the links that
our method identifies incorrectly, we provide evidence as to possible
reasons.  Furthermore, we are able to identify many of the key nonlinear relations
among variables in the model.

In the absence of good approximation methods over finite fields, our method does exact fitting
of data and is consequently quite sensitive to noise.  Future work includes the development of
approximation methods that will alleviate some of this sensitivity.  A more systematic study of
different data discretization methods and their effect on noise reduction will also be carried out.
Finally, the method will be validated using both simulated networks and published models that
generate real-valued data.

%%%%%%%%%%%%%%%%%%%%%%%%%%%%%%%%%%%%%%%%
\section{Acknowledgements}
The research in this paper was partly supported by NSF grants
DMS-0138323 and DMS0083595, and NIH grant RO1GM068947-01.  The authors thank J. Shah for the implementation of
the algorithms discussed in this paper, and O. Col\'on-Reyes, A. de la Fuente, L. Garcia, E. Green, A. Li, P. Mendes,
E. Nordberg, J. Snoep, M. Stillman, and B. Sturmfels
for helpful discussions.  Finally, the authors thank the anonymous referees for their helpful comments.

%%%%%%%%%%%%%%%%%%%%%%%%%%%%%%%%%%%%%%%%%%%%%%%%%%%%%%%%%
% Bibliographic references with the natbib package:
% Parenthetical: \citep{Bai92} produces (Bailyn 1992).
% Textual: \citet{Bai95} produces Bailyn et al. (1995).
% An affix and part of a reference:
%   \citep[e.g.][Ch. 2]{Bar76}
%   produces (e.g. Barnes et al. 1976, Ch. 2).
% \bibitem[Names(Year)]{label} or \bibitem[Names(Year)Long names]{label}.
% (\harvarditem{Name}{Year}{label} is also supported.)
% Text of bibliographic item
%\bibitem[]{}

%%%%%%%%%%%%%%%%%%%%%%%%%%%%%%%%%%%%%%%%%%%%%%%%%%%%%%%%
\pagebreak
% The Appendices part is started with the command \appendix;
% appendix sections are then done as normal sections
\appendix
% \section{}
% \label{}
\textbf{Appendix}
\section{Mathematical Background}
\label{math}
In this section we give the mathematical details of the reverse-engineering algorithm and
some basic facts about computational algebra relevant to this paper.  The basic problem
that lies at the heart of computational algebra is that long division of multivariate polynomials
differs from that of univariate polynomials in that the remainder is not uniquely determined.
It depends on several choices that need to be made along the way.   Univariate division
of a polynomial $f$ by another one, $g$, proceeds by dividing the highest power of the
variable in $g$ into the highest power in $f$.  For multivariate polynomials there are 
many choices of ordering the terms of a polynomial, which affects the outcome of the
division.  Also, when dividing more than one polynomial into a given one, the outcome
typically depends on the order in which the division is carried out.  To be precise,
let $k$ be a field, e.g. the field of real or complex numbers, or a finite field, and let
$f,f_1,\ldots, f_m\in k[x_1,\ldots ,x_n]$ be polynomials in the variables $x_1,\ldots ,x_n$.
The question whether there are polynomials $g_1,\ldots ,g_m\in k[x_1,\ldots ,x_n]$
such that
$$
f=\sum_{i=1}^mg_if_i
$$
can in general not be answered algorithmically.  In the language of abstract algebra,
let $I=\langle f_1,\ldots ,f_m\rangle$ be the {\it ideal} in $k[x_1,\ldots ,x_n]$ generated
by the $f_i$, that is, $I$ is the set of all linear combinations $\sum g_if_i$, 
with $g_i\in k[x_1,\ldots ,x_n]$.  Then we are asking whether $f$ is an element of $I$.
This is known as the {\it ideal membership problem}.  

In general, the ideal $I$
can be generated by sets of polynomials other than the $f_i$, similar to 
a vector space possessing many different bases, and the solution to the ideal membership
problem does not depend on the choice of a particular generating set.  It turns out that
if one chooses a very special type of generating set, known as a {\it Gr\"obner basis},
for the ideal, then the ideal membership problem becomes solvable algorithmically.
There is a basic algorithm to compute a Gr\"obner basis for an ideal, the 
Buchberger algorithm.  Using this algorithm as a foundation, many problems in
multivariate computational algebra can be solved algorithmically, including the
computation of intersections of ideals, which is the key computation we are using
in our algorithm.  The central ingredient in this algorithm is the 
Elimination Theorem (see (Cox \textit{et al.}, 1997), p. 113).  This theorem is also the source of
the very high computational complexity of many algorithms, since it requires the
use of a computationally expensive type of term ordering.  For details see, e.g., 
(Cox \textit{et al.}, 1997).  

As described earlier, the basic idea of our algorithm is as follows.  First we choose
a term order for $k[x_1,\ldots ,x_n]$.  This is necessary for all subsequent calculations.
To compute the ideal $I$ of all polynomial functions that are identically equal to $0$
on a given collection $\{\bf s_i\}$ of points we proceed as follows.
Let $I_i$ be the ideal of all polynomials that take on the value $0$ on $\mathbf s_i$.
It is straightforward to see that $I_i$ contains the ideal
$$
\langle x_1-s_{i1},\ldots ,x_1-s_{in}\rangle .
$$
But this ideal is maximal with respect to inclusion, so that it has to be equal to $I_i$.
Then the ideal $I$ is equal to the intersection 
$$
I=\bigcap_{i=1}^mI_i.
$$
Algorithms to compute the intersection of ideals are implemented in many specialized
computer algebra systems.  As mentioned earlier, we have used {\it Macaulay2} for all
computations.
Finally the reduction of the special solution
$f_0$ modulo $I$ is simply the remainder of $f_0$ under division
by a reduced Gr\"obner basis of $I$.  This too is a standard algorithm implemented in
most computer algebra systems.  It is important to observe that, if $g_0$ is another
particular solution to the interpolation problem, then $f_0$ and $g_0$ differ by a
polynomial in $I$, as shown before.  Therefore, the reduction of $f_0$ by $I$ is
equal to the reduction of $g_0$.  Consequently, it does not matter which method
we use to construct a particular solution.

%%%%%%%%%%%%%%%%%%%%%%%%%%%%%%%%%%%%%
\section{Tables}
\label{tables}

\begin{table}[h]
\begin{tabular}{|l l l|}
\hline
$f_{1}$ &=&$SLP_{i}^{t+1}=\left\{ 
\begin{array}{ccccc}
0 & if &i \, mod \, 4=1 & or & i \, mod \, 4=2 \\ 
1 & if &i \, mod \, 4=3 & or & i \, mod \, 4=0
\end{array}
\right. $ \\
$f_{2}$ &=&$wg_{i}^{t+1}=\left( CIA_{i}^{t}\wedge SLP_{i}^{t}\wedge \lnot
CIR_{i}^{t}\right) \vee \left( wg_{i}^{t}\wedge \left( CIA_{i}^{t}\vee
SLP_{i}^{t}\right) \wedge \lnot CIR_{i}^{t}\right) $ \\
$f_{3}$ &=&$WG_{i}^{t+1}=wg_{i}^{t}$ \\
$f_{4}$ &=&$en_{i}^{t+1}=\left( WG_{i-1}^{t}\vee WG_{i+1}^{t}\right) \wedge
\lnot SLP_{i}^{t}$ \\
$f_{5}$ &=&$EN_{i}^{t+1}=en_{i}^{t}$ \\
$f_{6}$ &=&$hh_{i}^{t+1}=EN_{i}^{t}\wedge \lnot CIR_{i}^{t}$ \\
$f_{7}$ &=&$HH_{i}^{t+1}=hh_{i}^{t}$ \\
$f_{8}$ &=&$ptc_{i}^{t+1}=CIA_{i}^{t+1}\wedge \lnot EN_{i}^{t+1}\wedge \lnot
CIR_{i}^{t+1}$ \\
$f_{9}$ &=&$PTC_{i}^{t+1}=ptc_{i}^{t}\vee \left( PTC_{i}^{t}\wedge \lnot
HH_{i-1}^{t}\wedge \lnot HH_{i+1}^{t}\right) $ \\
$f_{10}$ &=&$PH_{i}^{t+1}=PTC_{i}^{t+1}\wedge \left( HH_{i-1}^{t+1}\vee
HH_{i+1}^{t+1}\right) $ \\
$f_{11}$ &=&$SMO_{i}^{t+1}=\lnot PTC_{i}^{t+1}\vee HH_{i-1}^{t+1}\vee
HH_{i+1}^{t+1}$ \\
$f_{12}$ &=&$ci_{i}^{t+1}=\lnot EN_{i}^{t}$ \\
$f_{13}$ &=&$CI_{i}^{t+1}=ci_{i}^{t}$ \\
$f_{14}$ &=&$CIA_{i}^{t+1}=CI_{i}^{t}\wedge \left( SMO_{i}^{t}\vee
hh_{i-1}^{t}\vee hh_{i+1}^{t}\right) $ \\
$f_{15}$ &=&$CIR_{i}^{t+1}=CI_{i}^{t}\wedge \lnot SMO_{i}^{t}\wedge \lnot
hh_{i-1}^{t}\wedge \lnot hh_{i+1}^{t}$ \\ 
\hline
\end{tabular}
\smallskip
\caption{Boolean functions for the network in Figure \ref{bool}.}
\label{bf}
\end{table}

\begin{table}[h]
\begin{tabular}{|l l l|}
\hline
$f_{1}$ &=&$x_{1} $\\
$f_{2}$ &=&$x_{1}x_{14}\left( x_{15}+1\right) +x_{2}\left( x_{1}+x_{14} +x_{1}x_{14}\right) \left( x_{15}+1\right) +x_{1}x_{2}x_{14}\left( x_{15}+1\right) ^2\left( x_{1}+x_{14}+x_{1}x_{14}\right) $\\
$f_{3}$ &=&$x_{2} $\\
$f_{4}$ &=&$\left( x_{16}+x_{17}+x_{16}x_{17}\right) \left( x_{1}+1\right) $\\
$f_{5}$ &=&$x_{4} $\\
$f_{6}$ &=&$x_{5}\left( x_{15}+1\right)  $\\
$f_{7}$ &=&$x_{6} $\\
$f_{8}$ &=&$x_{13}\left( \left( x_{11}+x_{20}+x_{11}x_{20}\right) +x_{21}+ \left( x_{11}+x_{20}+x_{11}x_{20}\right) x_{21}\right)\left( x_{4}+1\right) $\\ 
&& $\left( x_{13}\left( x_{11}+1\right) \left( x_{20}+1\right) \left( x_{21}+1\right) +1\right)  $\\
$f_{9}$ &=&$x_{8}+x_{9}\left( x_{18}+1\right) \left( x_{19}+1\right)
+x_{8}x_{9}\left( x_{18}+1\right) \left( x_{19}+1\right)  $\\
$f_{10}$ &=&$\left( x_{8}+x_{9}\left( x_{18}+1\right) \left( x_{19}+1\right)
+x_{8}x_{9}\left( x_{18}+1\right) \left( x_{19}+1\right) \right) \left(
x_{20}+x_{21}+x_{20}x_{21}\right)  $\\
$f_{11}$ &=&$x_{8}+x_{9}Y+x_{8}x_{9}Y +1+x_{20}+ \left( \left( x_{8}+x_{9}Y+x_{8}x_{9}Y +1\right)
x_{20}\right) +x_{21} $\\ 
&& $+\left( x_{8}+x_{9}Y+x_{8}x_{9}Y +1+x_{20}+ \left( x_{8}+x_{9}Y +x_{8}x_{9}Y +1\right) x_{20}\right) x_{21} $\\ 
%&& $(Y:=\left( x_{18}+1\right) \left( x_{19}+1\right) ) $\\
$f_{12}$ &=&$x_{5}+1 $\\
$f_{13}$ &=&$x_{12} $\\
$f_{14}$ &=&$x_{13}\left( \left( x_{11}+x_{20}+x_{11}x_{20}\right)
+x_{21}+\left( x_{11}+x_{20}+x_{11}x_{20}\right) x_{21}\right)  $\\
$f_{15}$ &=&$x_{13}\left( x_{11}+1\right) \left( x_{20}+1\right) \left(
x_{21}+1\right) $\\
\hline
\end{tabular}
%\bigskip
\begin{tabular}{|c|c|c|c|c|c|c|c|c|c|c|} 
\hline
SLP & $wg$ & WG & $en$ & EN & $hh$ & HH & $ptc$ & PTC & PH & SMO \\ \hline
$x_{1}$ & $x_{2}$ & $x_{3}$ & $x_{4}$ & $x_{5}$ & $x_{6}$ & $x_{7}$ & $x_{8}$ & $x_{9}$ & $x_{10}$ & $x_{11}$ \\ \hline
$ci$ & CI & CIA & CIR & WG$_{i-1}$ & WG$_{i+1}$ & HH$_{i-1}$ & HH$_{i+1}$ & $hh_{i-1}$ & $hh_{i+1}$ & $Y$ \\ \hline
$x_{12}$ & $x_{13}$ & $x_{14}$ & $x_{15}$ &
$x_{16}$ & $x_{17}$ & $x_{18}$ & $x_{19}$ & $x_{20}$ & $x_{21}$ & $\left( x_{18}+1\right) \left( x_{19}+1\right) $ \\ \hline
\end{tabular}
\smallskip
\caption{Polynomial representations of the Boolean functions in Table \ref{bf}, together 
with the legend of variable names.}
\label{pf}
\end{table}

\begin{table}[h]
\begin{tabular}{|l l l|}
\hline
$f_{1} $&=&$x_{1} $\\
$f_{2} $&=&$x_{1}x_{14}+x_{2}x_{14}+x_{2}x_{15}+x_{2} $\\
$f_{3} $&=&$x_{2} $\\
$f_{4} $&=&$x_{16} $\\
$f_{5} $&=&$x_{4} $\\
$f_{6} $&=&$x_{5} $\\
$f_{7} $&=&$x_{6} $\\
$f_{8} $&=&$x_{12}x_{13}+x_{13}x_{16}+x_{13}x_{20}+x_{18}x_{20}+x_{13}x_{21}+x_{19}x_{21}+x_{13}+x_{18}+x_{19} $\\
$f_{9} $&=&$x_{10}x_{14}+x_{14}x_{18}+x_{14}x_{19}+x_{9}x_{20}+x_{18}x_{20}+x_{9}x_{21} $\\ 
&& $+x_{19}x_{21}+x_{8}+x_{9}+x_{10} $\\
$f_{10} $&=&$x_{10}x_{14}+x_{14}x_{18}+x_{14}x_{19}+x_{8}x_{20}++x_{8}x_{21}+x_{10}+x_{18}+x_{19} $\\
$f_{11} $&=&$ x_{8}x_{20}+x_{9}x_{20}+x_{18}x_{20}+x_{8}x_{21}+x_{9}x_{21}+x_{19}x_{21}+x_{8}+x_{9}+x_{18}+x_{19}+1
$\\
$f_{12} $&=&$x_{5}+1 $\\
$f_{13} $&=&$x_{12} $\\
$f_{14} $&=&$x_{11}x_{13}+x_{9}x_{20}+x_{18}x_{20}+x_{9}x_{21}+x_{19}x_{21}+x_{10}+x_{18}+x_{19} $\\
$f_{15} $&=&$x_{11}x_{13}+x_{9}x_{20}+x_{18}x_{20}+x_{9}x_{21}+x_{19}x_{21}+x_{10}+x_{13}+x_{18}+x_{19} $\\
\hline
\end{tabular}
\smallskip
\caption{Boolean functions reduced by the ideal of wildtype and knock-out time series.}
\label{rbf}
\end{table}

%%%%%%%%%%%%%%%%%%%%%%%%%%%%%%%%%%%%%%%

\begin{thebibliography}{12}

\bibitem{akutsu}
Akutsu, T., Miyano, S., and Kuhara, S., 1999.  Identification of genetic networks from a small 
number of gene expression patterns under the Boolean network model.  In
Altman, R. B., Lauderdale, K., Dunker, A. K., Hunter, L., and Klein, T. E. (eds.),
{\it Proc. Pacific Symp. Biocomput.} {\bf 4}, 17--28, World Scientific, Singapore.

\bibitem{albert} Albert, R. Othmer, H., 2003. The topology of the regulatory interactions predicts the expression
pattern of the segment polarity genes in \textit{Drosophila melanogaster}. {\it J. Theor. Biol.} {\bf 223}, 1--18.

\bibitem{bower}
Bower, J. M., and Bolouri, H., 2001. {\it Computational Modeling of Genetic and Biochemical
Networks}, The MIT Press, Cambridge, MA.

\bibitem{brazma}
Brazma, A. and Schlitt, T., 2003.  Reverse engineering of gene regulatory networks: a 
finite state linear model.  preprint, available at {\tt http://genomebiology.com/2003/4/6/P5}.

\bibitem{cadigan}  Cadigan, K.M., Grossniklaus, U., Gehring, W.J., 1994. Localized
expression of \textit{sloppy paired} protein maintains the polarity of
\textit{Drosophila} parasegments. {\it Genes Dev.} {\bf 8}, 899--913.

\bibitem{chen}
Chen, Y., and Struhl, G., 1998.  In vivo evidence that Patched and Smoothened constitute distinct binding
and transducing components of a Hedgehog receptor complex. \textit{Development} \textbf{125}, 4943--4948.

\bibitem{cho}
Cho, R., Campbell, M., Winzeler, E., Steinmetz, L., Conway, A., Wodicka, L., Wolfsberg, T.,
Gabrielan, A., Landsman, D., Lockhart, D., and Davis, R., 1998.  A genome-wide
transcriptional analysis of the mitotic cell cycle.  {\it Mol. Cell} {\bf 2}, 65--73.

\bibitem{cox} Cox, D., Little, J., and O'Shea, D., 1997.  {\it Ideals, Varieties, and Algorithms},
Springer Verlag, New York.

\bibitem{delafuente}
de la Fuente, A., Brazhnik, P., and Mendes, P., 2002.  Linking the genes: inferring quantitative
gene networks from microarray data.  {\it Trends in Genetics} {\bf 18}, 395--398.

\bibitem{dejong}
de Jong, H., 2002.  Modeling and simulation of genetic regulatory systems: a literature review.
{\it J. Comp. Biol.} {\bf 9}, 67--103.

\bibitem{dougherty}
Dougherty, J., Kohavi, R., and Sahami, M., 1995.  Supervised and unsupervised discretization 
of continuous features.  In Prieditis, A. and Russell, S. (eds.), {\it Machine learning: Proceedings
of the 12th International Conference}, Morgan Kauffman, San Francisco, CA.

\bibitem{filkov-a}
Filkov, V., Skiena, S., and Zhi, J., 2002.  Analysis techniques for microarray time-series data.
{\it J. Comp. Biol.} {\bf 9}, 317--330.

\bibitem{filkov-b}
Filkov V., and Istrail, S., 2002.  Inferring gene transcription networks: the Davidson model.
{\it Genome Informatics} {\bf 13}, 236--239.

\bibitem{friedman}
Friedman N., Linia, l. N. M., Nachman, I., and Pe'er D., 2000.  Using Bayesian
networks to analyze expression data, {\it J. Comp. Biol.} {\bf 7}, 601--620.

\bibitem{friedman2}
Friedman, N., and Goldszmidt, M., 1996.  Discretization of continuous attributes while
learning Bayesian networks.  In Saitta, L. (ed.), {\it Proc. of the 13th International Conference on
Machine Learning}, 157--165, Morgan Kauffman, San Francisco, CA.

\bibitem{green}
Green, E. 2003.  On polynomial solutions to reverse-engineering problems.  Preprint.

\bibitem{hartemink}
Hartemink, A. J., Gifford, D. K., Jaakkola S., and Young, R. A., 2001.
Using graphical models and genomic expression data to statistically validate
models of genetic regulatory networks.  {\it Pac. Symp. Biocomput.}, World Scientific,
Singapore.

\bibitem{hooper}  Hooper, J.E., Scott, M.P., 1992. The molecular genetic basis of
positional information in insect segments, in: Hennig, W. (Ed.),
{\it Early Embryonic Development of Animals}, Springer, Berlin, 1–-49.

\bibitem{ingham-a}
Ingham, P. W., 1998. Transducing hedgehog: the story so far. {\it EMBO J.} 1{\bf 7}, 3505–-3511.

\bibitem{ingham-b}
Ingham, P. W., McMahon, A. P., 2001. Hedgehog signaling in
animal development: paradigms and principles. {\it Genes Dev.} {\it 15},
3059–3087.

\bibitem{kauffman}
Kauffman, S. A., 1969.  Metabolic stability and epigenesis in randomly constructed
genetic nets.  {\it J. Theor. Biol.} {\bf 22}, 437--467.

\bibitem{kitano}
Kitano, H., 2002. Systems biology: A brief overview.  {\it Science} {\bf 295}, 1662--1664.

\bibitem{lang} 
Lang, S., 1971.  {\it Algebra}.  Addison Wesley, Reading, MA.

\bibitem{lewis}
Lewis, J.E., and Glass, L., 1991.  Steady states, limit cycles, and chaos in models
of complex biological networks.  {\it Int. J. Bifurcation and Chaos} {\bf 1}, 477--483.

\bibitem{liang}
Liang, S., Fuhrman, S., and Somogyi, R., 1998.  REVEAL, a general reverse engineering
algorithm for inference of genetic network architectures.  {\it Pac. Symp. Biocomput.} {\bf 3},
18--29.  World Scientific, Singapore.

\bibitem{lidl}
Lidl, R., and Niederreiter, H., 1997.  \emph{Finite Fields}, Encyclopedia of
Mathematics and its Applications 20, 2nd Edition, Cambridge
University Press.

\bibitem{marchand}
Marchand, H., and Le Borgne, M., 1998.  Partial order control of discrete event systems modeled
as polynomial dynamical systems.  in {\it IEEE International Conference on Control Applications},
Trieste, Italy.

\bibitem{mendes}
Mendes, P., Sha, W., and Ye, K., 2003. Artificial gene networks for objective comparison
of analysis algorithms. {\it Bioinformatics} {\bf 19}, 122--129.

\bibitem{mendoza}
Mendoza, L., Thieffry, D., and Alvarez-Buylla, E. R., 1999.  Genetic control of flower morphogenesis
in {\it Arabidopsis thaliana}: a logical analysis.  {\it Bioinformatics} {\bf 15}, 593--606.

%\bibitem{pfeiffer}  Pfeiffer, S., and Vincent, J. P., 1999. Signalling at a distance: 
%transport of Wingless in the embryonic epidermis of \textit{Drosophila}. {\it Semin. Cell Dev. Biol.} {\bf 10}, 303--309.

\bibitem{repsilber}
Repsilber, D., Liljenstr\"om, H., and Andersson, S. G. E., 2002.  Reverse engineering of regulatory
networks: simulation studies on a genetic algorithm approach for ranking hypotheses.
{\it BioSystems} {\bf 66}, 31--41.

\bibitem{robbiano} Robbiano L., 1998.  Gr\"obner bases and statistics, in {\it Gr\"obner Bases and Applications},
Buchberger B. and Winkler F. (eds.), Cambridge University Press, New York.

\bibitem{shmulevich}
Schmulevich, I., Dougherty, E. R., Kim, S., and Zhang, W., 2002. Probabilistic Boolean networks: a rule-based
uncertainty model for gene regulatory networks. {\it Bioinformatics} {\bf 18}, 261--274.

\bibitem{shparlinski} Shparlinski, I. E., 1999. {\it Finite Fields: Theory and Computation}.  Kluwer Academic
Publishers, Dortrecht.

\bibitem{spellman}
Spellman, P. T., Sherlock, G., Zhang, M. Q., Iyer, V. R., Anders, K., Eisen, M. B., Brown, P. O., Botstein, D.,
and Futcher, B., 1998.  Comprehensive identification of cell cycle-regulated genes of the yeast
{\it Saccharomyces cerevisiae} by microarray hybridization.  {\it Mol. Biol. Cell} {\bf 9}, 3273--3297.

\bibitem{taylor}  Taylor, A.M., Nakano, Y., Mohler, J., Ingham, P.W., 1993.
Contrasting distributions of patched and hedgehog proteins in the \textit{Drosophila} embryo. 
{\it Mech. Dev.} {\bf 42}, 89--96.

\bibitem{thieffry1}
Thieffry, D., and Thomas, R., 1998.  Qualitative analysis of gene networks.  {\it Proc. Pacific Symp. on Biocomputing},
77--88, World Scientific, Singapore.

\bibitem{thieffry}
Thieffry, D., Thomas, R., and Kaufman, M., 1995.  Dynamical behaviour of biological regulatory networks--I. 
Biological role of feedback loops and practical use of the concept of the loop-characteristic state.
{\it Bulletin of Math. Bio.} {\bf 57}, 247--276.

\bibitem{thomas}
Thomas, R., 1991.  Regulatory networks seen as asynchronous automata: a logical description.
{\it J. Theor. Biol.} {\bf 153}, 1--23.


\bibitem{von}  von Dassow, G., Meir, E., Munro, E. M., Odell, G. M., 2000. The segment polarity network is a robust developmental module. {\it Nature} {\bf 406}, 188--192.

\bibitem{yeung}
Yeung. M. K. S., Tegn\'er, J., and Collins, J. J., 2002.  Reverse engineering gene networks using singular
value decomposition and robust regression.  {\it Proc. Natl. Acad. Sci.} {\bf 99}, 6163--6168.

\end{thebibliography}
\end{document}